\documentclass[aps,twocolumn]{revtex4}
\usepackage{amssymb}
\usepackage{amsmath}
\usepackage{graphicx}
\usepackage{epsfig,amsmath}
\usepackage[bookmarksnumbered,linktocpage,pdfstartview=FitH]{hyperref}

\begin{document}

\title{On the Nonuniform Quantum Turbulence in Superfluids}
\author{Sergey K. Nemirovskii\thanks{%
email address: nemir@itp.nsc.ru}}
\affiliation{Institute of Thermophysics, Lavrentyev ave, 1, 630090, Novosibirsk, Russia\\
and Novosibirsk State University, Novosibirsk}
\date{\today }

\begin{abstract}
The problem of quantum turbulence in a channel with an inhomogeneous
counterflow of superfluid turbulent helium is studied. \ The counterflow
velocity $V_{ns}^{x}(y)$ along the channel is supposed to have a parabolic
profile in the transverse direction $y$. Such statement corresponds to the
recent numerical simulation by Khomenko et al. [Phys. Rev. B \textbf{91},
180504 (2015)]. The authors reported about a sophisticated behavior of the
vortex line density (VLD) $\mathcal{L}(\mathbf{r},t)$, different from $%
\mathcal{L}\propto V_{ns}^{x}(y)^{2}$, which follows from the naive,
straightforward application of the conventional Vinen theory. It is clear,
that Vinen theory should be refined by taking into account transverse
effects and the way it ought to be done is the subject of active discussion
in the literature. In the work we discuss several possible mechanisms of the
transverse flux of VLD $\mathcal{L}(\mathbf{r},t)$ which should be
incorporated in the standard Vinen equation to describe adequately the
inhomogeneous quantum turbulence (QT). It is shown that the most effective
among these mechanisms is the one that is related to the phase slippage
phenomenon. The use of this flux in the modernized Vinen equation corrects
the situation with an unusual distribution of the vortex line density, and
satisfactory describes the behavior $\mathcal{L}(\mathbf{r},t)$ both in
stationary and nonstationary situations. The general problem of the
phenomenological Vinen theory in the case of nonuniform and nonstationary
quantum turbulence is thoroughly discussed.
\end{abstract}

\maketitle

\section{Introduction.}

The question of evolution of the vortex line density (VLD) $\mathcal{L}(%
\mathbf{r},t)$ of the vortex tangle (VT) is the key issue in the macroscopic
theory of quantum turbulence (QT). Although the VLD is a rough
characteristic of the QT, it is responsible for many (mainly hydrodynamic)
phenomena in superfluids and the knowledge of its exact dynamics is very
important for an adequate interpretation of various experiments.

Long ago Vinen \cite{Vinen1957c} suggested that the rate of change of VLD $%
\partial \mathcal{L}(t)/\partial t$\ can be described in terms of only the
quantity $\mathcal{L}(t)$\ itself (and also other, external parameters, such
as the counterflow velocity $V_{ns}$\ and the temperature). He called this
statement as a self-preservation assumption. The corresponding balance
equation for the quantity $\mathcal{L}(r,t)$, the so called Vinen equation,\
reads:
\begin{equation}
\frac{\partial \mathcal{L}}{\partial t}=~\alpha _{V}\;|\mathbf{v}_{ns}|\;%
\mathcal{L}^{3/2}\;-\;\beta _{V}\;\mathcal{L}^{2}.  \label{VE 1}
\end{equation}%
Here\textit{\ }$~\alpha _{V}\;$and$\;\beta _{V}\;$are the parameters of the
theory, $\alpha _{V}$ is close to the mutual friction coefficient $\alpha $,
$\beta _{V}$ is of the order of the quantum of circulation $\kappa $.
Throughout its long history, the Vinen equation has undergone various
improvements and modifications (see e.g. \cite{Schwarz1988}, \cite%
{Nemirovskii1995},\cite{Jou2011},\cite{Donnelly1991},\cite{Nemirovskii2013}
although at present the form (\ref{VE 1}) is mainly used.

One of serious problems, is the application of the Vinen theory to
complicated situations, in particular to inhomogeneous flows (for recent
papers see e.g. \cite{Khomenko2015},\cite{Baggaley2015}, \cite{Yui2015},
\cite{Saluto2016}, \cite{Kivotides2011}). In the cited papers the authors,
analyzing numerically the steady counterflowing helium in an inhomogeneous
channel flow, obtained a very specific behavior of the VLD $\mathcal{L}(%
\mathbf{r},t)$, which cannot be interpreted in terms of equation (\ref{VE 1}%
). Thus, Khomenko et al. \cite{Khomenko2015} observed that the VLD field is
concentrated near the side walls. Quite similar behavior was observed in the
work by Yui et al. \cite{Yui2015}.

Analyzing the obtained results, the authors of the paper \cite{Khomenko2015}
proposed, that the first term on the right hand side of the Vinen equation
(the so called production term) has the structure $\propto \left\vert
\mathbf{V}_{ns}\right\vert ^{3}\mathcal{L}^{1/2}$, a combination that has
never been discussed before. This conclusion was the subject of a polemics
between the authors of the article \cite{Khomenko2015} and the author of
this paper (see \cite{Nemirovskii2016PRB} and \cite{Khomenko2016}).

In the present paper, I would like to digress from the content of the
mentioned polemics, and to present our view on the macroscopic behavior of
the VLD\ in inhomogeneous flows (referring to numerical results of the work
\cite{Khomenko2015}). In the study I retain the conventional form of the
production term in the Vinen equation (\ref{VE 1}).

In short, the results of work \cite{Khomenko2015} can be formulated as
follows. In a rectangular channel $2\times 0.05$ cm wide, a parabolic
counterflow $V_{ns}^{x}(y)=V_{0}(1-(y/0.05)^{2})$ is applied in $x$
direction. The periodic conditions were assumed in all directions. The
resulting distributions of the dimensional VLD, the normal and counterflow
velocities $\mathcal{L}(y),V_{n}(y),V_{ns}(y)$ are presented in Fig. \ref%
{Khomenko1}.

If one applies straightforwardly the well-known relation $\mathcal{L=}\gamma
^{2}V_{ns}^{x}(y)^{2}\approx 2\ast 10^{4}$ $V_{ns}^{x}(y)^{2}$ (here $\gamma
=\alpha _{V}/\beta _{V}$), which immediately arises from equation (\ref{VE 1}%
), then the\ dimensionless $\mathcal{L}$ should be about $2\ast 10^{-2}$ $%
V_{ns}^{x}(y))^{2}$, which essentially exceeds the value obtained in \cite%
{Khomenko2015}. Another striking feature is that the profile $\mathcal{L}(y)$
is radically different from the quadratic velocity profile $\mathcal{%
L\propto (}V_{ns}^{x}(y))^{2}$.

\begin{figure}[tbp]
\includegraphics[width=7cm]{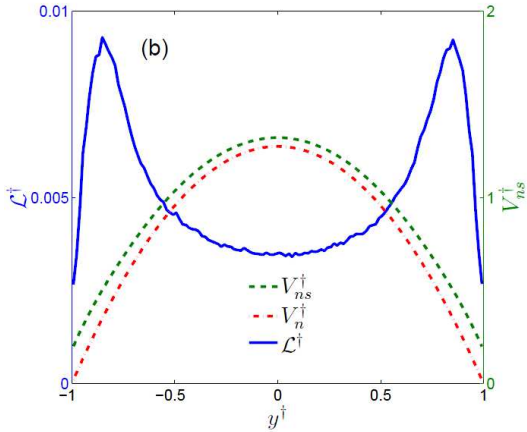}
\caption{(Color online) Prescribed parabolic normal velocity profile $V_{n}$
(-- $\cdot $ --), the resulting counterflow profile $V_{ns}$ (-- --) and the
resulting profile of $\mathcal{L}(y)$ (--) in dimensionless unites, T = 1.6
K. (from paper by Khomenko et al. \protect\cite{Khomenko2015})}
\label{Khomenko1}
\end{figure}

In the paper we develop an approach explaining this unusual (from the point
of view of the naive use of the Vinen theory) behavior of the VLD\ $\mathcal{%
L}(y)$. In the inhomogeneous situation the Vinen equation should be
corrected to include the transverse spacial effects. In particular we offer
to incorporate into classic Vinen theory an additional space flux $\mathbf{J}%
(\mathbf{r},t)$ of the VLD, which redistributes the quantity $\mathcal{L}(y)$
in the $y$ direction. It is clear that a transverse\ gradient of the flux%
\emph{\ }$\partial J_{y}(y,t)/\partial y$\emph{\ }should be added into the
balance equation (\ref{VE 1}).

In the next Sec. II\ we discuss several mechanisms of these possible fluxes,
derive mathematical expressions and compare contributions from them. In Sec.
III we present numerical solutions for stationary and nonstationary cases
and compare the results with the numerical data of paper \cite{Khomenko2015}%
. In Sec. IV we discuss the problem of nonuniform and unsteady quantum
turbulence and the Vinen phenomenological theory. The Conclusion is devoted
to a discussion of the results and probable generalizations of the presented
approach.

\section{Vortex-line density flux}

\label{flux}

Let's describe various ideas on the transverse\ vortex-line density flux%
\emph{\ }$\mathbf{J}(\mathbf{r},t)$ in inhomogeneous flows/counterflows of
superfluid helium. As it was mentioned above, the first remark in this
respect had been made by Vinen himself in the context of the possible
influence of the channel width \cite{Vinen1957c}. Unfortunately, no advanced
theory had been supplemented. It is clear that the most general expression
for the flux of quantity $\mathcal{L}$ is $\mathbf{J}(\mathbf{r},t)=\mathcal{%
L}\mathbf{V}_{L}$, where $\mathbf{V}_{L}$ is the macroscopic local velocity
of the vortex tangle (see explanations in papers\ \cite{Nemirovskii1983},%
\cite{Geurst1989},\cite{Nemirovskii1995}). However, unless we don't have a
general expression for $\mathbf{V}_{L}$ as a function (functional) of
quantity $\mathcal{L}$, we can not ascertain a closure procedure, i.e.
obtain a description of the vortex tangle dynamics in terms of the VLD
itself. This procedure is not uniquely defined and admits different
approaches.

Thus, in the cited paper \cite{Khomenko2015} \ the authors proceeded from
the following microscopic expression for the transverse flux $J_{micro}$

\begin{equation}
J_{micro}=\frac{1}{\Omega }\int \left\vert \mathbf{V}_{ns}(y)\right\vert
\mathbf{s}_{z}^{\prime }d\xi =\frac{\alpha }{\Omega }\int \left\vert \mathbf{%
V}_{ns}(y)\right\vert \mathbf{s}_{z}^{\prime }d\xi .  \label{Khom micro}
\end{equation}%
Here the integration is performed over the whole vortex line configuration,
so it should be understood as an integration along each vortex loops
constituting the vortex tangle and summation over all loops, i.e.
\begin{equation*}
\int d\xi \rightarrow \sum_{j}\int\limits_{0}^{L_{j}}d\xi _{j}.
\end{equation*}%
The quantity $\Omega $ is the total volume, $\alpha $ is the mutual friction
coefficient. The authors of work \cite{Khomenko2015} calculated the quantity
(\ref{Khom micro}) in numerical simulation and concluded that the
macroscopic expression%
\begin{equation}
\mathbf{J}_{Kh}(\mathbf{r},t)=\frac{\alpha }{2\kappa }C_{flux}\frac{\partial
\mathbf{V}_{ns}^{2}}{\partial y},  \label{Jkh}
\end{equation}%
best corresponds to the microscopic flux (\ref{Khom micro}). The quantity $%
C_{flux}$ is a constant, determined from numerical simulations. Another
mechanism, frequently discussed in the problems of nonuniform flow, is
related to the diffusion flux \cite{Geurst1989},\cite{Tsubota2003a}. That
mechanism is not connected with mutual friction, and realized by the
emission of vortex loops, (see, e.g., \cite{Barenghi2002},\cite%
{Kondaurova2012}). The diffusion flux can be written as follows

\begin{equation}
\mathbf{J}_{dif}(\mathbf{r},t)=D\mathbf{\nabla }\mathcal{L},  \label{Jdif}
\end{equation}%
where the diffusion coefficient is estimated as $D\approx 2\ast 10^{-3}$ cm$%
^{2}$/s.

The next contribution, which we consider here, is related to the so called
phase slippage phenomenon. This phenomenon implies appearance of additional
the chemical potential $\nabla \mu $, and accordingly the mutual friction
when the crossing by the vortices of the main flow. This effect is
especially important for monitoring the quantization of vortices. We will
use the corresponding technique to describe the transverse flux of VLD $%
J_{y}(y,t)$. To find an analytical expression for $J_{y}(y,t)$, consider the
following equation (see \cite{Rasetti1975}, \cite{Nemirovskii1998}, \cite%
{Swanson1985})%
\begin{equation}
\mathcal{A}\ =\int \left( {\mathbf{\dot{s}}}(\xi )\times \mathbf{s}^{\prime
}(\xi )\right) d\xi .  \label{sheet}
\end{equation}

The right-hand side of (\ref{sheet}) is a net area, swept out by the motion
of the line elements. Therefore, the $x$ -component of vector $\mathcal{A}$
is simply the rate of phase slippage (without the factor $\kappa $) caused
by the transverse motion of the vortex lines (see \cite{Swanson1985}). It is
important, however, that the sign of the $x$ -component of the vector $%
\mathcal{A}$ does not depend on the direction of motion of vortex line
segments (either in the positive or in the negative directions along axis $y$%
). It makes no differences in the calculation of the phase slippage, and
accordingly the additional drop in the chemical potential $\nabla \mu $, but
it is essential for our purposes to determine flux $J_{ps}(y,t)$ of the VLD $%
\mathcal{L}$ to the side wall. To overcome this problem we assume that all
the vortex filaments are closed loops, so the averaged fluxes in both
directions are equal. Therefore, the required transverse flux $J_{ps}(y,t)$
of the VLD $\mathcal{L}$ is just half of the $x$ -component of the vector $%
\mathcal{A}$. Taking velocity of elements ${\mathbf{\dot{s}}}(\xi )$ in the
form \ of the local induction approximation (see e.g. \cite{Schwarz1988}),
we arrive at the following expression

\begin{equation}
J_{ps}(\mathbf{r},t)=\frac{1}{2}\int \left( \left[ \alpha \mathbf{s}^{\prime
}\times \left( \mathbf{V}_{ns}-\beta (\mathbf{s}^{\prime }\times \mathbf{s}%
^{\prime \prime })\right) \right] \times \mathbf{s}^{\prime }(\xi )\right)
d\xi .  \label{Ax}
\end{equation}

Here the combination $\dot{\mathbf{s}}_{i}=\beta (\mathbf{s}^{\prime }\times
\mathbf{s}^{\prime \prime })$ is the self-induced velocity of the line
elements in the the local induction approximation.

To move further we have to introduce the closure procedure and to express
the right hand side of Eq. (\ref{Ax}) via quantities $\mathcal{L}$ and $%
\mathbf{V}_{ns}$. It corresponds to the self-preservation assumption
expressed by Vinen, that the macroscopic dynamics of the vortex tangle
depends only on the VLD ${\mathcal{L}}(t)$. The other, more subtle
characteristics of the vortex structure, different from ${\mathcal{L}}$,
must adjust to it. In particular, the first contribution, containing the
external counterflow velocity can be written as $\alpha I_{\parallel }%
\mathcal{L}\left\vert \mathbf{V}_{ns}\right\vert $, where $I_{\parallel }$
is the structure parameter of the vortex tangle, introduced by Schwarz \cite%
{Schwarz1988}. The last term in Eq. (\ref{Ax}) with the self-induced
velocity can be expressed as $\alpha \beta \mathcal{L}(I_{l}\mathcal{L}%
^{1/2})$. where $I_{l}$ is another structure parameter. Usually at this
point the substitution $\mathcal{L}^{1/2}=\gamma \left\vert \mathbf{V}%
_{ns}\right\vert $ is used, and both contributions are reduced to a
combination
\begin{equation}
J_{ps,1}(\mathbf{r},t)=\frac{1}{2}\alpha (I_{\parallel }-\gamma \beta I_{l})%
\mathcal{L}\left\vert \mathbf{V}_{ns}\right\vert .  \label{J1}
\end{equation}

Being multiplied by $\rho _{s}\kappa $ this expression (up to a factor $1/2$%
) coincides with the formula for mutual friction. This is not surprising,
because it is well known from the vortex dynamics that a vortex crossing the
channel transfers the momentum to the main flow (see \cite{Nemirovskii1998b}%
). Therefore the final expression should be proportional to $\mathbf{V}_{ns}$
and the whole scheme becomes self-consistent. But this above consideration
concerns only homogeneous or near - homogeneous cases. In the highly
inhomogeneous situation, which we are interested in here, the simple
relations such as $\mathcal{L}^{1/2}=\gamma \left\vert \mathbf{V}%
_{ns}\right\vert $ do not work and the question of determining the
transverse flux remains open. A very similar problem of using the structure
parameters of the vortex tangle also arises for nonstationary situations
(see a related discussion in the review article \cite{Nemirovskii1995}).
This problem is very intriguing, and we decided to explore yet another
version of the closure procedure, which leads to the following formula for
the transverse flux
\begin{equation}
J_{ps}(y,t)=\alpha I_{\parallel }\mathcal{L}\left\vert \mathbf{V}%
_{ns}\right\vert -\alpha \beta I_{l}\mathcal{L}^{3/2}.  \label{J2}
\end{equation}

Thus, we have obtained two forms for the transverse flux associated with the
phase slippage mechanism. They are identical in case of an uniform flow ,
when $\mathcal{L}^{1/2}=\gamma \left\vert \mathbf{V}_{ns}\right\vert $,
however, in inhomogeneous situations they differ and can result in different
results.

Our further goal is to analyze the results on the nonuniform quantum
turbulence obtained in\ the numerical\ work by Khomenko et al. \cite%
{Khomenko2015}, basing on supposition of the transverse\ flux of VLD $%
\mathcal{L}(y)$. \ Using the conditions of their modeling and taking that $%
\left\vert \mathbf{V}_{ns}\right\vert \sim 1$ cm/s, $\mathcal{L}\sim 10^{4}$
1/cm$^{2}$, $\alpha \sim 0.1$, $\partial /\partial y\sim 1/0.05$, we
conclude that the most effective mechanism among those considered above, is
the one related to the phase slippage mechanism. It exceeds other
contributions almost by the order and further we will concentrate on the
only this effect.

Beside the usual estimation and comparison of various fluxes written above
we can appeal to the fact that neither Khomenko et al. flux $\mathbf{J}%
_{Kh,cl}$ no the diffusion flux $\mathbf{J}_{dif}$ are effective enough to
produce the complicated spacial distribution of the vortex line density
which was observed in paper \cite{Khomenko2015} and is shown in Fig \ \ref%
{Khomenko1}. As far as the Khomenko et al. flux $\mathbf{J}_{Kh,cl}$\ this
problem was discussed in details in the paper \cite{Nemirovskii2016PRB} \
(Sec. IV).

The impact of the diffusion flux was studied in a recent work by Saluto et
al. \cite{Saluto2016}. The authors observed that the influence of the vortex
diffusion is focused on the local values of $\mathcal{L}(y)$ rather than on
the form of the spatial distribution VLD. Thus the diffusion term (\ref{Jdif}%
) is also small for this particular problem, although, being a second-order
derivative, it would be essential for other situations. In this paper we
will not consider this term.

\section{ Solutions}

Thus we introduced and discussed several mechanisms for the transverse flux
of VLD and concluded that the most effective of them is associated with the
phase slippage mechanism. A microscopic equation for this flux is given by
Eq. ( \ref{Ax}), its macroscopic closure variants are given by the formulas (%
\ref{J1}),(\ref{J2}). Our goal now is to incorporate these terms into the
Vinen equation (\ref{VE 1})
\begin{equation}
\frac{\partial \mathcal{L}}{\partial t}+\frac{\partial J_{ps}(y,t)}{\partial
y{}}=~\alpha _{V}\;|\mathbf{V}_{ns}|\;\mathcal{L}^{3/2}\;-\;\beta _{V}\;%
\mathcal{L}^{2},  \label{VE flux}
\end{equation}%
and to study its solutions under the conditions that identical to those
studied in the work by Khomenko et al. \cite{Khomenko2015}. Namely, we have
selected the temperature of system, the geometry and size of the of the
channel, parabolic counterflow velocity $\mathbf{V}_{ns}(y)$ coinciding with
the ones accepted in their work. We study two cases, a stationary situation
and a completely unsteady problem.

\subsection{Stationary case, profile of VLD $\mathcal{L}(y)$.}

\begin{figure}[tbp]
\includegraphics[width=7cm]{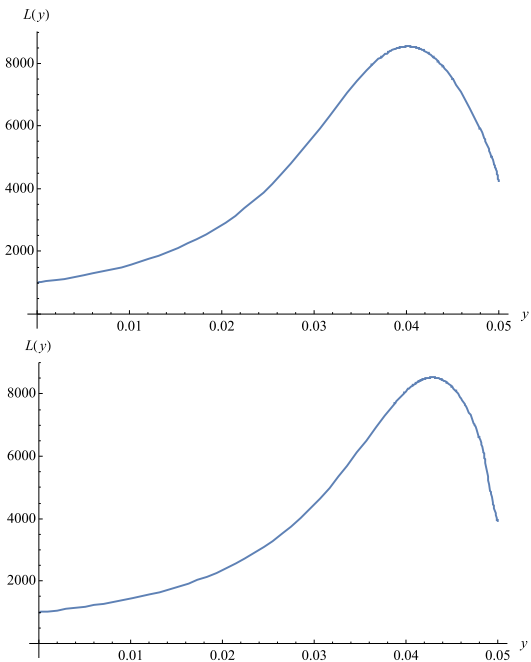}
\caption{(Color online) Profiles of VLD $\mathcal{L}(y)$ obtained in
numerical solution of the equation (\protect\ref{VE flux}) without the term $%
\partial \mathcal{L}/\partial t$. The upper and lower pictures correspond to
different expressions for transverse flux (\protect\ref{J1}),(\protect\ref%
{J2}). }
\label{stationary}
\end{figure}

In Fig. \ref{stationary} we displayed the VLD $\mathcal{L}(y)$ profiles
obtained through the numerical solution of equation (\ref{VE flux}) without
the term $\partial \mathcal{L}/\partial t$. The upper and lower images
correspond to different expressions for the transverse flux (\ref{J1}),(\ref%
{J2}). We have chosen the system temperature $T=1.6$ K, the channel size $%
2\ast 0.05$ cm, the parabolic counterflow velocity $\mathbf{V}_{ns}(y)$ $%
=1.2(1-(y/0.05)^{2})$ cm/s, coinciding with the conditions adopted in the
work \cite{Khomenko2015}. Additionally, only half of the channel width is
considered, namely $0\eqslantless y\eqslantless 0.05$ cm. The boundary
condition $\mathcal{L}(y=0)=1000$ 1/cm$^{2}$ had been taken from the result
of paper \cite{Khomenko2015} and from the solution of the fully
nonstationary problem (see below). It is noteworthy that they are very close
to each other.

The most important (albeit expected) result is that the VLD profile does not
really satisfy the standard Vinen relation $\mathcal{L}(y)=\gamma ^{2}|%
\mathbf{V}_{ns}|^{2}$. On the contrary, the vortex tangle is concentrated in
the region closer to the side wall, (but not directly on the wall). This
behavior can be understood qualitatively from the following considerations.
The structure of flux expressed by the formula (\ref{J1}) is that its
maximal value is at the central parts ($y=0$) of the channel (due to the
large value of the counterflow velocity $V_{ns}$) and the VLD $\mathcal{L}$
is intensively removed from this region. On the contrary, because of the
vanishing of the counterflow velocity $V_{ns}$ on the side walls ($y=0.05$),
the flux is almost extinguished, and $\mathcal{L}$ does not penetrate into
this region. Clearly, to support a stationary solution in the regions where $%
\mathcal{L}(y)\neq \gamma ^{2}|\mathbf{V}_{ns}|^{2}$, either the production
or the decay (second) term on the right hand side of equation (\ref{VE flux}%
) should prevail. Another remarkable result is that there is a very good
agreement, both qualitative and quantitative, with the data of the paper
\cite{Khomenko2015} depicted in Fig. \ref{Khomenko1}.

One more important result concerns the fundamental question of the use of
the Schwarz's relations for the structure parameters of the nonuniform
quantum turbulence \ In the lower picture of Fig. \ref{stationary} we
presented the quantity $\mathcal{L}(y)$ obtained in numerical solution of
the equation (\ref{VE flux}) with the transverse flux expressed by Eq. (\ref%
{J2}), which includes an alternative variant of the structure parameter. \
It is easy to see that qualitatively solutions are very similar, although
they are a bit different. This fact confirms the widespread view that the
Vinen equation can be a good tool for studying rough engineering problems,
although relevant approaches may require some fitting parameters. At the
same time the whole Vinen macroscopic theory is not suitable for the
investigation of the fine structure of the vortex tangle.

\subsection{Nonstationary case, development of quantum turbulence in the
inhomogenious counterflow.}

The rather elegant results are obtained when solving the full equation (\ref%
{VE flux}), with the term $\partial \mathcal{L}/\partial t$. This procedure
faces the standard problem of initial conditions, typical for the Vinen
theory. Equation ((\ref{VE flux})) is a balance relation between the growth
and the disappearance of vortex lines. The mechanism of spontaneous
appearance of vortices in the helium flow has not been built into this
equation.

At present, there are various theories of the initial appearance of vortex
filament, which can be divided into two groups. The first group offers the
different mechanisms (tunnelling, fluctuation growth, etc.) of initial
generation of vortices. Another group is based on the idea that in the
helium permanently\ exists a background of remnant vortices. From the point
of view of the phenomenological theory the former group can be taken into
account by introducing the initiating term into the Vinen equation. In turn,
the latter group should lead to some initial value of VLD ($\mathcal{L}(t=0)=%
\mathcal{L}_{back}$) in the Vinen equation. The better agreement between
experimental data on the propagation of intense heat pulses (generating
vortices and interacting with these "own" vortices) and the corresponding
numerical solution, was obtained when assuming the existence of an initial
level of VLD $\mathcal{L}_{back}$, whereas the introduction of the
initiating term led to an unsatisfactory correlation with the experimental
observations (see e.g. \cite{Kondaurova2017}). Thus it may be surmised that
this is an argument in favour of the theory of remnant vortices. Usually,
the level of the remnant vorticity $\mathcal{L}_{back}\ $is estimated
approximately as $10^{2}-10^{3}$ 1/cm$^{2}$.
\begin{figure}[tbp]
\includegraphics[width=8cm]{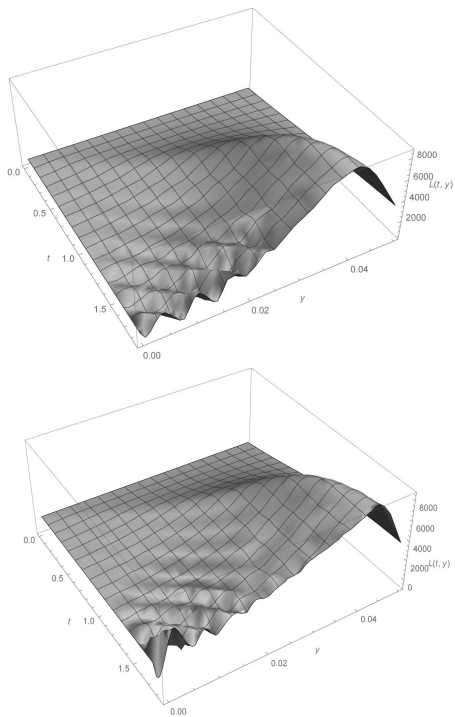}
\caption{(Color online) The spatio - temporal behavior of VLD $\mathcal{L}%
(t,y)$ obtained in numerical solution of the equation (\protect\ref{VE flux}%
).The upper and lower pictures correspond to different expressions for
transverse flux (\protect\ref{J1}),(\protect\ref{J2}). }
\label{nonstationary}
\end{figure}

The spatio - temporal behavior of VLD $\mathcal{L}(y,t)$ obtained in the
numerical solution of the equation (\ref{VE flux}) with the nonstationary
term $\partial \mathcal{L}/\partial t$ is shown in Fig. \ref{nonstationary}.
The upper and lower images the correspond to the different expressions for
the flux (\ref{J1}),(\ref{J2}). We again have chosen all conditions of work
\cite{Khomenko2015}. As for initial conditions we assume that the background
vorticity $\mathcal{L}_{back}=1000$ 1/cm$^{2}$. The obtained picture
confirms all the conclusions on the behavior of the VLD\ $\mathcal{L}(y,t)$,
made in the previous paragraph, and demonstrates how the according scenario
is developing in time. On a time slice of $2$ s (It is probable saturation
and crossover to the steady-state regime), the solution $\mathcal{L}(y,t=2\
c)$ agrees with the data found in Ref. \cite{Khomenko2015} (see also Fig. %
\ref{Khomenko1}). That is a remarkable fact because in our study no fitting
parameters have been used.

\section{Nonuniform quantum turbulence and the Vinen phenomenological theory}

In Sec. \ref{flux} \ we described the problems of the closure procedure for
the microscopic equation for the flux ( \ref{Ax}) and questions of the
choice of the form for the structure parameters. Bearing in mind to compare
various possibilities we have chosen two variants, leading to different
expressions (\ref{J1}),(\ref{J2}). In this regard, it seems appropriate to
return to the basics of Vinen's phenomenological theory as applied to the
complex nonstationary and inhomogenious situations.

The main idea of the Vinen approach was the assumption of self-preservation
, i.e. the suggestion that the macroscopic vortex dynamics can be described
in terms of the quantity $\mathcal{L}(t)$ only. Selecting a set of variables
to describe the macroscopic dynamics of statistical systems is, in general,
a difficult and delicate step. For instance, the usual gas dynamics
variables, such as density, momentum and energy (per unit volume) are just
the first moments of the distribution function of the Boltzmann's kinetic
theory. Higher moments relax to approach equilibrium much faster than do the
first listed variables. This circumstance allows one to truncate an infinite
hierarchy of the moment equations and obtain a closed description using the
listed quantities.

Unfortunately in case quantum turbulence, the assumption of
self-preservation is not motivated, the restriction to the only variable ${%
\mathcal{L}}(t)$ is not justified, and, in general, the Vinen equation is
not valid. Indeed, let us consider a very simple counterexample. Assume that
the velocity $\mathbf{V}_{ns}(\mathbf{s},t)$ changes instantly to the
opposite. Since the Vinen-type equation include the absolute value of
relative velocity $\left\vert \mathbf{V}_{ns}(\mathbf{s},t)\right\vert $
magnitude, then formally the system remains unaffected by the change. This
is wrong, of course. The structure of the VT, mean curvature, anisotropy and
polarization parameters will become reorganized. That implies the violation
of the self-preservation assumption, and dynamics of the VLD ${\mathcal{L}}%
(t)$ depends on other, more subtle characteristics of the vortex structure,
different from ${\mathcal{L}}(t)$.

To clarify the situation, let us consider a way of derivation of VE from the
dynamics of vortex filaments in the local induction approximation (see, e.g.
\cite{Schwarz1978}). It will suffice for the illustration sake. Integrating
an equation for the change of the length of line element over $\xi $ inside
a volume $\Omega $,\ Schwarz concluded that in the counterflowing helium II
\ the quantity ${\mathcal{L}}(t)$ obeys the equation (see \cite{Schwarz1988}%
)
\begin{equation}
\frac{\partial \mathcal{L}}{\partial t}=\;\frac{\alpha \mathbf{V}_{ns}}{%
\Omega }\int \left\langle \mathbf{s}^{\prime }\times \mathbf{s}^{\prime
\prime }\right\rangle \;d\xi \;-\frac{{\alpha \beta }}{\Omega }\int
\left\langle |\mathbf{s}^{\prime \prime }|^{2}\right\rangle \;d\xi \;.
\label{VLD rate}
\end{equation}%
The quantity ${\mathcal{L}}(t)$ is related to the first derivative $\mathbf{s%
}^{\prime }$ of the function $\mathbf{s}(\xi )$, since ${\mathcal{L}}%
(t)\propto \int |\mathbf{s}^{\prime }|d\xi $. The rate of change of ${%
\mathcal{L}}(t)$ includes quantities involving the higher-order derivative $%
\mathbf{s}^{\prime \prime }$, namely $\left\langle \mathbf{s}^{\prime
}\times \mathbf{s}^{\prime \prime }\right\rangle $ and $\left\langle |%
\mathbf{s}^{\prime \prime }|^{2}\right\rangle $. In a steady-state, these
higher-order quantities are are directly expressed via the VLD $\mathcal{L}$
as $\left\langle \mathbf{s}^{\prime }\times \mathbf{s}^{\prime \prime
}\right\rangle \propto $ $I_{l}\mathcal{L}^{1/2\text{ }}$and $\left\langle |%
\mathbf{s}^{\prime \prime }|^{2}\right\rangle \propto c_{2}^{2}(T)\mathcal{L}
$. Here the $I_{l},c_{2}(T)$ are temperature dependent parameters introduced
by Schwarz \cite{Schwarz1988}. But in the nonstationary situation $\mathbf{s}%
^{\prime \prime }$ is a new independent variable, and one needs a new
independent equation for it and for other quantities, related to curvature
of line. This new equation, in turn, will involve higher derivatives $%
\mathbf{s}^{\prime \prime \prime },\mathbf{s}^{IV}$ and so on. This infinite
hierarchy can be truncated if, for some reasons, the higher-order
derivatives relax faster, than the low-order derivatives, and take their
"equilibrium" values (with respect to the moments of low order).

Strictly speaking, there are no theoretical grounds for \ assuming that the
relaxation of higher moments is faster than that of the quantity ${\mathcal{L%
}}(t)$. Thus, in general, no equation of the type{\Large \ }$\partial {%
\mathcal{L}}(t)/\partial t=\mathcal{F}({\mathcal{L}})${\Large \ }exists! At
the same time, in some (unclear) conditions, and with the use of additional
arguments (see, \cite{Vinen1957c}), the required equation can be written
down. The attempt was successful, this theory explained a large number of
hydrodynamic experiments, including the main experiment by Gorter and
Mellink \cite{Gorter1949} (see, for details, the review by \cite{Tough1982}%
). It concerned , however, only stationary or near-stationary situations. In
a strongly unsteady case, the region of applicability of this equation is
unclear\emph{, }see the above counterexample with a sudden inversion of the
counterflow velocity.

Meanwhile, it seems intuitively plausible that for slow changes (both in
space and time) the assumption of self-preservation is valid. That was the
starting point in the construction of the so-called Hydrodynamics of
Superfluid Turbulence (HST), which was the unification of the Vinen equation
and the classical two-fluid hydrodynamics (see, e.g., \cite{Nemirovskii1983},%
\cite{Yamada1989},\cite{Geurst1992}). The HST\ equations have been applied
to study a large number of hydrodynamic and thermal problems, including heat
transfer and boiling in He II (see, e.g., \cite{Fiszdon1990},\cite%
{Murakami1990}, \cite{Murakami1991},\cite{Poppe1992}, \cite{Tsoi1986},\cite%
{Ruppert1987},\cite{Kondaurova2017}). The numerical and analytic results
were in very good agreement with numerous experimental data. This fact
pointed out that the Vinen equation is robust and is, in general, quite
suitable for the unsteady hydrodynamic problems.

It follows from the results of this work that the situation with
inhomogenious flow is quite similar. This is confirmed by the curves
depicted on the upper and lower images in Figures \ref{stationary}\ and \ref%
{nonstationary}. In these images we display the results obtained from
solutions of the Vinen equation ( \ref{VE flux}) with different expressions (%
\ref{J1}),(\ref{J2}) for the transverse flux. The qualitative similarity and
closeness of the quantitative solutions indicates again that the Vinen
equation is rather insensitive to a particular choice of the transverse flux
and\ is robust to study various inhomogenious situations.

\section{Conclusion}

We conclude by saying that the study of the inhomogenious flow/counterflow
of superfluids in the channel on the basis of the Vinen equation (\ref{VE 1}%
) requires the introduction of additional terms describing the transverse
flux of the VLD $\mathcal{L}$ towards the side walls. The analysis
demonstrated that the most efficient mechanism is related to the phase
slippage mechanism. The corresponding solutions of the Vinen equation with
the additional term in both stationary and nonstationary cases agree with
observations obtained earlier in numerical simulations. They showed that the
VLD $\mathcal{L}(y,t)$, as function of $y$ is concentrated in the domain
near the side walls. The reason for this behavior is the special structure
of the transverse flux. This construction forces the vortex filaments to
escape from the central part, at the same time does not allow them to touch
the walls.

One of our results, important for the macroscopic theory of quantum
turbulence concerns the structure functions of the vortex tangle, such as
the parameters of anisotropy and polarization. Just like in the unsteady
situation, the use of such parameters in the usual form, introduced by
Schwarz, can only be done approximately and with reservations. This fact
confirms the widespread view that the Vinen equation can be used to explore
the rough, engineering problems (although the corresponding studies may
require some fitting parameters), but it's not suitable for the description
of the fine structure of the vortex tangle.

I would like to thank Prof. I. Procaccia for the very fruitful discussion of
questions touched in the paper. The work was supported by Grant No.
14-29-00093 from RSCF (Russian Scientific Foundation)


\end{document}